\documentstyle[preprint,prl,aps]{revtex}

\tightenlines 
\narrowtext
\draft

\begin{document}
\title{On the $T-$dependence of the magnetic penetration depth in unconventional
superconductors at low temperatures: can it be linear?}
\author{N. Schopohl and O.V. Dolgov}
\address{Eberhardt-Karls Universit\"{a}t T\"{u}bingen\\
Institut f\"{u}r Theoretische Physik\\
Auf der Morgenstelle 14\\
D-72076 T\"{u}bingen\\
Germany}
\maketitle

\begin{abstract}
We present a thermodynamics argument against a strictly linear temperature
dependence of the magnetic penetration depth, which applies to
superconductors with arbitrary pairing symmetry at low temperatures.
\end{abstract}


Some evidence for an unconventional $d_{x^{2}-y^{2}}-$pairing symmetry in
cuprate high$-T_{c\text{ }}$superconductors is provided by recent angle
resolved photoemission experiments\cite{photoemission}. A striking proof for
the $d_{x^{2}-y^{2}}-$symmetry of the Cooper pairs in cuprates arises from
the observation of a spontanously generated {\em half flux quantum} in
Josephson tunneling experiments carried out on tetracrystal substrates\cite
{half-flux}. Early support for the possibility of a $d_{x^{2}-y^{2}}-$%
symmetry of the Cooper pairs in cuprate high$-T_{c\text{ }}$superconductors
came from the observation of a {\em linear }$T-$dependence of the magnetic
penetration depth \cite{exp1},\cite{exp2} at low temperatures $T$: 
\begin{equation}
\lambda (T)-\lambda (0)\sim T
\end{equation}
\tightenlines command somewhere after the documentstyle command Such a
linear $T-$dependence of the magnetic penetration depth (MPD) has a
topological origin. If the order parameter associated with the Cooper pair
condensate vanishes along node lines on the Fermi surface the spectrum $%
N_{s}(E)$ of quasiparticle excitations in the superconducting phase is
gapless and varies proportional to $E$ at low excitation energies: $%
N_{s}(E)\sim E$ for $E\ll \Delta _{\max }$ . For this reason a pure $%
d_{x^{2}-y^{2}}-$pairing state (node lines along $k_{x}=\pm k_{y}$) should
display a strictly {\em linear} dependence of MPD vs. $T$ at {\em low}
temperatures. In previous work this effect was also discussed for the {\em %
polar} phase in a triplet pairing superconductor, e.g. \cite{p-p}.

New experiments\cite{experiments} indicate deviations from this linearity of
MPD with temperature, for example a $T^{2}$-dependence of MPD below some
cross over temperature $T^{*}$ was measured. Such a behaviour may occur due
to various reasons. For example, I. Kosztin and A.J. Leggett \cite{legg}
explaine this behavior in terms of {\em nonlocal} electrodynamics. Their
argument is, that in clean $d_{x^{2}-y^{2}}-$pairing superconductors there
exist {\em surface induced} non local effects, which lead to a $T^{2}$%
-dependence of $\lambda _{ab}(T)-\lambda _{ab}(0)$, as extracted from
optical and microwave experiments with the magnetic field orientated {\em %
parallel} to the $\widehat{c}-$ direction. On the other hand, in experiments
with the magnetic field orientated {\em perpendicular} to the $\widehat{c}-$
direction the $T$-dependence of MPD cannot be altered by the Kosztin-Leggett
effect.

Since the Kosztin-Leggett effect \cite{legg} really depends on the existence
of a {\em surface} in the problem it cannot be applied to other measurement
techniques of MPD, for example direct static magnetic measurements,
measurements of vortex properties, the lower critical magnetic field $B_{c1}$%
, muon spin relaxation. Such techniques of measuring MPD have {\em bulk}
character.

In the following we present a proof, for arbitrary superconductors, that a
strictly linear $T-$dependence of MPD at low temperatures violates the third
law of thermodynamics. For simplicity let us consider a uniform system where
all properties depend on coordinates ${\bf r-r}^{\prime }$ only. The
current-current correlator, 
\begin{equation}
\eta ({\bf k,}\omega )=k^{2}-\frac{\omega ^{2}}{c^{2}}\varepsilon _{tr}({\bf %
k,}\omega )
\end{equation}
connects the vector potential ${\bf A(k,}\omega )$ to the external current $%
{\bf j}_{ext}{\bf (k,}\omega )$ via

\begin{equation}
\eta ({\bf k,}\omega ){\bf A(k\,,}\omega )=\frac{4\pi }{c}{\bf j}_{ext}{\bf %
(k,}\omega )
\end{equation}
\vspace*{10pt} In turn, the transversal dielectric function, $\varepsilon
_{tr}({\bf k,}\omega )$, is related to the electromagnetic kernel $Q({\bf k,}%
\omega )$ by the relation 
\begin{equation}
\varepsilon _{tr}({\bf k,}\omega )=1-\frac{4\pi Q({\bf k,}\omega )}{\omega
^{2}}
\end{equation}
The definition of the operator of inverse MPD is then 
\begin{equation}
\frac{1}{\lambda ^{2}({\bf k,}T)}=\lim_{\omega \rightarrow 0}\frac{\omega
^{2}}{c^{2}}\{1-%
\mathop{\rm Re}%
\varepsilon _{tr}({\bf k,}\omega )\}\equiv \frac{4\pi }{c^{2}}Q({\bf k,}%
\omega =0)
\end{equation}
In the {\em static} case the additional free energy in the presence of an
externaly controlled current distribution {\bf \ }${\bf j}_{ext}({\bf k})$
(we use a transversal gauge) can be written in the form \cite{we}:

\begin{eqnarray}
{\cal F} &=&-\frac{1}{2c}\int \frac{d^{3}k}{(2\pi )^{3}}{\bf j}_{ext}{\bf (k}%
)\cdot {\bf A(-k,}\omega =0) \\
&=&-\frac{1}{8\pi }\int \frac{d^{3}k}{(2\pi )^{3}}\ \eta ({\bf k,}\omega
=0)\ |{\bf A(k,}\omega =0)|^{2}  \nonumber
\end{eqnarray}
By using these relations and Maxwell's equations it follows

\begin{equation}
{\cal F}=-\frac{1}{8\pi }\int \frac{d^{3}k}{(2\pi )^{3}}\ \left[ k^{2}+\frac{%
1}{\lambda ^{2}({\bf k,}T)}\right] \ \frac{|{\bf k\times B(k\,};T)|^{2}}{%
k^{4}}  \label{free-energy}
\end{equation}
Here ${\bf B(k,}T)$ is the (temperature dependent) induced magnetic field
and satisfies the equation:

\begin{equation}
\left[ k^{2}+\frac{1}{\lambda ^{2}({\bf k,}T)}\right] {\bf B(k\,;}T)=\frac{%
4\pi }{c}\;i\,{\bf k\times j}_{ext}{\bf (k)}  \label{London}
\end{equation}
Differentiating Eq.(\ref{free-energy}) with respect to temperature $T$ and
calculating the derivative $\frac{\partial }{\partial T}{\bf B(k\,;}T)$ from
Eq.(\ref{London}) we get an expression for the entropy:

\begin{equation}
S\left( T\right) =-\frac{\partial {\cal F}}{\partial T}=-\frac{1}{8\pi }\int 
\frac{d^{3}k}{(2\pi )^{3}}\frac{\partial }{\partial T}\left[ \frac{1}{%
\lambda ^{2}({\bf k,}T)}\right] \frac{|{\bf B(k\,};T)|^{2}}{k^{2}}
\end{equation}
According to the Nernst principle ({\it third law of thermodynamics}) the
entropy should vanish in the limit $T\rightarrow 0$. From the positivity of
the integrand we must conclude

\begin{equation}
\lim_{T\rightarrow 0}\frac{\partial \lambda ({\bf k,}T{\bf )}}{\partial T}=0
\end{equation}
If we wish to avoid a violation of the third law of thermodynamics the $T-$%
dependence of the magnetic penetration depth in a superconductor {\em cannot}
be of the form $\lambda (T)-\lambda (0)\sim T^{n}$ with $n=1$. The argument
can be extended to any nonuniform system.

We see that the vanishing of the first derivative of MPD for $T\rightarrow 0$
is a consequence of a general principle of thermodynamics. The value of $%
T^{*}$ below which a deviation of the linear $T-$dependence of MPD may be
observed depends on the exact physical mechanism. It may be nonlocality\cite
{legg}, it may be the effect of impurities (as proposed in Ref.\cite{imp}),
it may be also the effect of collective excitations (e.g. the influence of
vertex corrections on the $T-$dependence of MPD was discussed in Ref.\cite
{eliash} for the case of pure $s-$wave pairing).

A famous reformulation of the third law of thermodynamics states that it is
impossible to reach absolute zero. From this point of view a pure $%
d_{x^{2}-y^{2}}-$pairing symmetry in clean high-$T_{c}$ superconductors
becomes, perhaps, invalid for $T\rightarrow 0$. A possibility to avoid the
paradox of a linear $T$-dependence of MPD for $T\rightarrow 0$ in cuprate
superconductors is a {\em phase transition} ( at a temperature $T_{c2}$ much
lower than the transition temperature $T_{c}$) to a new unconventional
pairing state {\em without} nodes on the Fermi surface\cite{loughlin}, \cite
{preosty}.

\medskip

{\bf Acknowledgements: }It is a pleasure to thank A.J. Leggett for helpful
correspondence and encouragement. Also we acknowledge useful discussions
with R.P. Huebener, D. Rainer, K. Scharnberg, C.C. Tsuei and G.E. Volovik.

\end{document}